\documentclass{article}

 \usepackage[final]{neurips_2025_ml4ps}

\usepackage[utf8]{inputenc} 
\usepackage[T1]{fontenc}    
\usepackage{hyperref}       
\usepackage{url}            
\usepackage{booktabs}       
\usepackage{amsfonts}       
\usepackage{nicefrac}       
\usepackage{microtype}      
\usepackage{xcolor}         

\usepackage{xspace}
\usepackage{svg}
\usepackage{tikz}
\usepackage{float} 

\newcommand{\chandra}{\textit{Chandra}\xspace}

\newcommand{\xmm}{\textit{XMM-Newton}\xspace}

\newcommand{\erosita}{\textit{eROSITA}\xspace}
\newcommand{\axis}{\textit{AXIS}\xspace}
\newcommand{\athena}{\textit{NewAthena}\xspace}

\title{Modeling X-ray photon pile-up with a normalizing flow}

\author{%
  Ole~K\"onig\\
  AstroAI, CfA $|$ Harvard \& Smithsonian\\
  \texttt{ole.koenig@cfa.harvard.edu} \And Daniela Huppenkothen\\ University of Amsterdam\\ \texttt{d.huppenkothen@uva.nl} \And Douglas Finkbeiner\\ Harvard University \& Anthropic\\ \texttt{dfinkbeiner@cfa.harvard.edu} \And Christian Kirsch\\ FAU Erlangen-N\"urnberg\\ \texttt{christian.ck.kirsch@fau.de} \And J\"orn Wilms\\ FAU Erlangen-N\"urnberg\\ \texttt{joern.wilms@fau.de} \And Justina R. Yang\\ Harvard University\\ \texttt{jyang1@g.harvard.edu} \And James F. Steiner\\ AstroAI, CfA $|$ Harvard \& Smithsonian\\ \texttt{james.steiner@cfa.harvard.edu} \And Juan Rafael Martínez-Galarza\\ AstroAI, CfA $|$ Harvard \& Smithsonian\\ \texttt{jmartine@cfa.harvard.edu} 
}

\begin{document}

\maketitle

\begin{abstract}
The dynamic range of imaging detectors flown on-board X-ray observatories often only covers a limited flux range of extrasolar X-ray sources. The analysis of bright X-ray sources is complicated by so-called pile-up, which results from high incident photon flux. This nonlinear effect distorts the measured spectrum, resulting in biases in the inferred physical parameters, and can even lead to a complete signal loss in extreme cases. Piled-up data are commonly discarded due to resulting intractability of the likelihood. As a result, a large number of archival observations remain underexplored. We present a machine learning solution to this problem, using a simulation-based inference framework that allows us to estimate posterior distributions of physical source parameters from piled-up \erosita data. We show that a normalizing flow produces better-constrained posterior densities than traditional mitigation techniques, as more data can be leveraged. We consider model- and calibration-dependent uncertainties and the applicability of such an algorithm to real data in the \erosita archive. 
\end{abstract}


\section{Introduction}

\begin{figure}[H]
    \centering
    \begin{tikzpicture}
        \node (p1) at (1.6,0){\includegraphics[height=2.7cm]{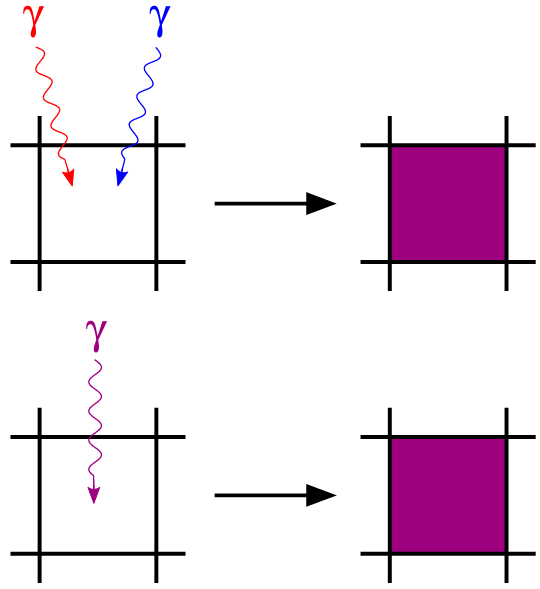}};
        \node (p2) at (-1.6,0){\includegraphics[height=2.7cm]{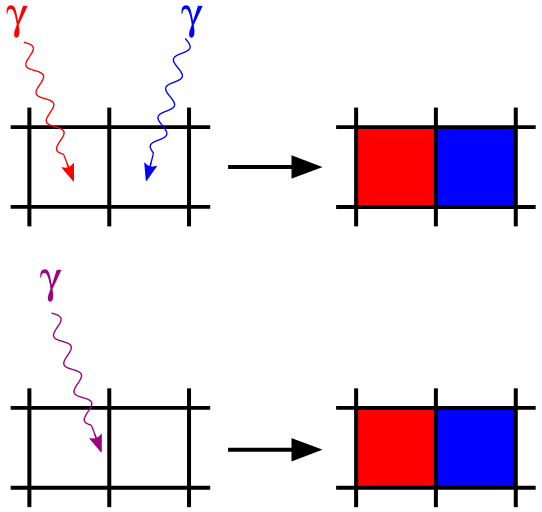}};
        \node (t1) at (p1.north){Energy pile-up};
        \node (t2) at (p2.north){Pattern pile-up};
    \end{tikzpicture}
    \hfill
    \begin{minipage}[b]{.54\linewidth}
        \caption{Illustration of X-ray photon pile-up in CCD detectors \citep{Schmid2012PhDT}. 
        \textit{Left:} A double event can be created either from one photon that lands near the border of the pixel (low photon flux) or two photons impacting both pixels (pattern pile-up). In the latter case, the event contains signal from the sum of the individual photons impacts (energy pile-up). \textit{Right:} Likewise, the same recorded single event can be created from one or two photons impacting the same pixel.}
        \label{fig:pileup_sketch}
    \end{minipage}
\end{figure}

In order to detect radiation from astrophysical X-ray sources, silicon-based charge-coupled devices (CCDs) have been used since 1993.
However, detectors flown on-board past, existing (\xmm, \chandra, \erosita) and future X-ray observatories (\athena/WFI, \axis) are unable to observe several classes of bright objects of interest without experiencing significant data distortion.
The analysis is complicated by pile-up, which is caused by high photon fluxes \citep{Ballet1999a,Davis2001a}. When an X-ray photon impacts a CCD, a cloud of electrons is created in the semiconductor \citep{Janesick1986a,PavlovNousek1999a}. These electrons are accumulated in one or multiple detector pixels, depending on the extent of the charge cloud. 
In the on-board event processing algorithm, each event is reconstructed from its associated charge distribution, and classified into four accepted ``patterns'': singles denote one pixel that contains charge during one readout cycle. If multiple adjacent pixels contain charge, the events are commonly classified into doubles (see colored boxes in Fig.~\ref{fig:pileup_sketch}), triples, and quadruples.

All currently operating X-ray CCD imaging telescopes operate in single-photon mode, which requires direct reconstruction of the energy of each incident photon. For a moderate photon flux, associating each pattern with a single photon is a valid assumption, and the original X-ray energy is inferred from the summed charges of the illuminated pixels. For sufficiently bright sources, however, multiple X-rays can hit the same or nearby pixels during one readout cycle (Fig.~\ref{fig:pileup_sketch}). In the extreme case, the illuminated pixels produce charge distributions that are different from the four accepted patterns, which is called ``pattern'' pile-up. This effect causes X-ray signal to be incorrectly rejected, leading to a loss of signal. 
At the core of the point spread function (PSF), where photon rates are highest, this can manifest as a flattening or even a depression in brightness (e.g., \citealt{Koenig2022a} and image in Fig.~\ref{fig:input_spectrum}). 
More commonly (i.e., at lower fluxes), the charge pattern deposited in the detector by multiple photons 
cannot be distinguished from that deposited by a single X-ray that has a higher energy, which is called energy pile-up \citep{Ballet1999a,Davis2001a}. As a consequence, the reconstructed count rate is reduced and the spectrum is artificially distorted toward higher energies (``hardened'').

The ramification of these effects on the data analysis can be severe. Pile-up is highly non-linear and estimating the source's photon flux and modeling the X-ray spectrum (i.e.~the source's brightness as a function of photon energy) without distortion is difficult. 
When pile-up becomes appreciable, model parameters obtained during parameter inference on X-ray spectra can be severely biased \citep[e.g.,][]{Dauser2019a,Koenig2022a}.
Ways to mitigate pile-up's biasing effects include analytical approaches \citep{Davis2001a,Jethwa2015a}, which do, however, not account for the loss of signal due to pattern pile-up, or using specialized detector read-out modes \citep{Duro2011a}. Another popular method is to excise the core of the PSF (e.g., \citealt{Malacaria2019a}), where pile-up is strongest due to the higher photon rates. This method discards data in the central source region (where most of the signal is concentrated) and uses only data from the wing of the PSF, where signals are much weaker. This leads to much wider posterior densities during inference, since the majority of signal power is discarded. Furthermore, the PSF excision radius is often determined ad hoc by the researcher, because it is often not obvious at which radius the strength of pile-up falls below an acceptable threshold.

While analytical models for pile-up exist only in very simplified cases, the stochastic nature of realistic pile-up prescriptions leads to intractable likelihoods, a circumstance well-matched to a simulation-based inference framework. Recent implicit likelihood analyses based on improved X-ray telescope simulators have used large grids of candidate models to match simulations to observed data based on simple test statistics \citep{Dauser2019a,Tamba2022a,Koenig2022a}. This method can use data from the whole source region but is computationally expensive and the researcher has to be trained on how to use the simulator.
Thus, due to the complex analysis techniques and computational burden, in practice, oftentimes, a piled-up dataset is dropped from the analysis, and all data is forfeit \citep[e.g.,][]{Steiner2010a}. 

\section{Methods and results}

We present a proof-of-concept neural network in a simulation-based inference (SBI) framework that approximates posterior probability distributions for the parameters of astrophysical models for X-ray spectra in the presence of pile-up. Specifically, the neural network takes spatially-resolved piled-up spectra as input, processes them through a convolutional neural network (CNN), and infers posteriors using a normalizing flow (NF). 

\paragraph{Input data}
Our SBI approach requires a forward model that takes physical parameters as input, simulates energy and pattern pile-up, and produces simulated data products analogous to the observed data. We use the \texttt{SIXTE} simulator \citep{Dauser2019a} to produce the piled-up spectra. To make the simulations representative of real \erosita data, we baseline the training data on an existing observation of a bright unresolved point source in the \erosita data archive, 
adopting the same source position and observing properties (nova V1710~Sco with a total exposure of 228\,seconds), although we note that the specific source choice does not influence the methodology development of this paper. We simulate 40\,000 \erosita observations using a model comprising a thermal blackbody source that varies in flux (on a logarithmic grid), temperature, and absorption (see Table~\ref{tab:ranges}), sampled using a Latin hypercube \citep{McKayBeckmanConover1979_latinhypercube}. 
Blackbody spectra are simple, easily interpretable, and have physical application, making them both practical and useful as a test case. Furthermore, there is a known parameter correlation between the absorption and temperature (a higher temperature can be compensated by lower absorption and vice versa), which the neural network, once trained on blackbody spectra, should be able to predict accurately.

\begin{figure}
    \centering
    \includegraphics[width=.8\linewidth]{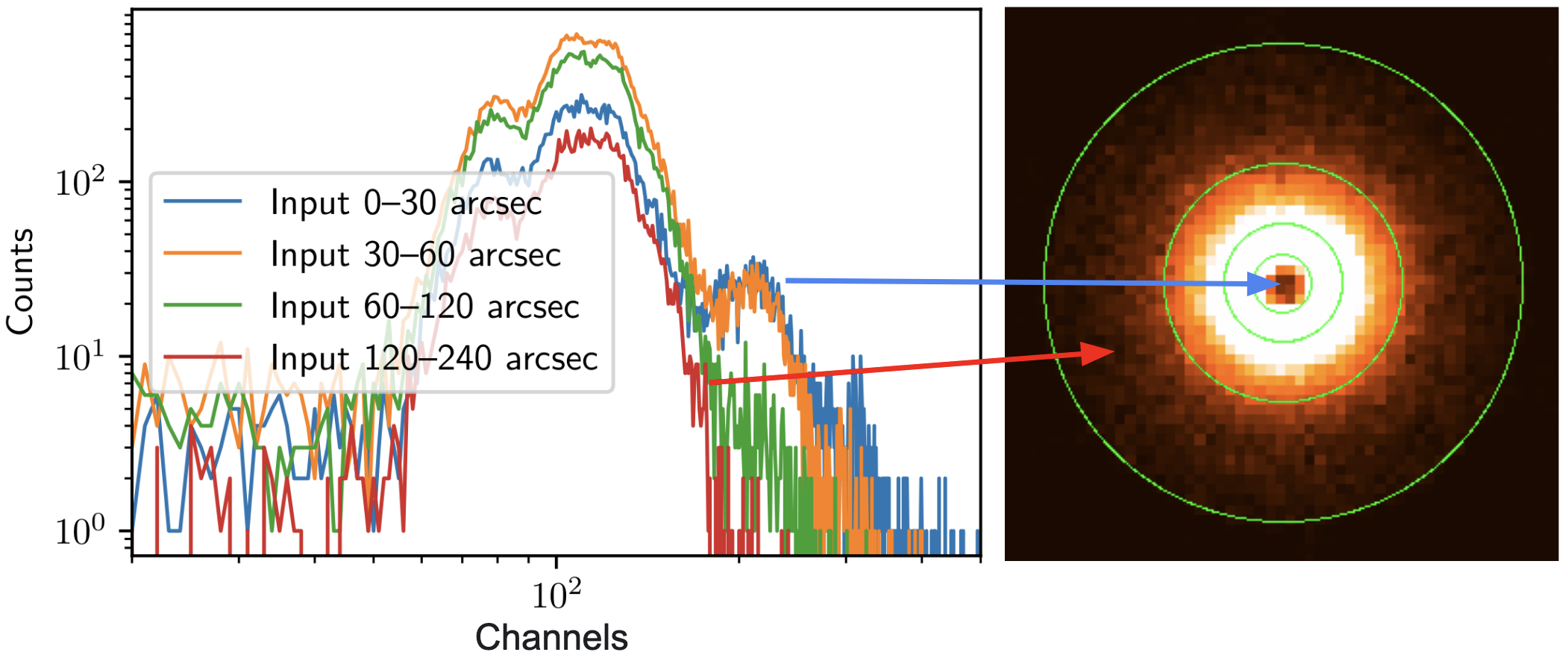}
    \caption{Example of a simulated piled-up \erosita observation in the training dataset, representing an  absorbed blackbody source with temperature of 130\,eV and absorption of $1.3\times 10^{22}\,\mathrm{cm}^{-2}$. Spectra of four annuli are used as input to the neural network. At a flux of $6.9\times 10^{-9}\,\mathrm{erg}\,\mathrm{cm}^{-2}\,\mathrm{s}^{-1}$, considerable pattern pile-up is present, clearly seen as a depression in brightness at the center of the image (blue arrow), as well as energy pile-up, which manifests as an artificial high-energy bump (blue and orange spectrum around channel 200). The outer annuli are much less affected by pile-up, as the photons are distributed over more pixels in the lower-intensity outer wing of the PSF.
    }
    \label{fig:input_spectrum}
\end{figure}

Pile-up distortion varies strongly over the detector, depending on how photons are spread over the pixels according to the PSF. At the center of the PSF with the highest impact rates, spectra suffer most pile-up (see blue spectrum in Fig.~\ref{fig:input_spectrum}). At outer radii, the PSF broadens and photons are spread over a larger area of the detector, creating less pile-up (red spectrum). This radial dependency contains information on how many photons impact the detector. A higher rate, for instance, leads to a larger radius up to which data loss occurs. Physically, this dependence primarily depends on the flux of the source. To take this information into account, for each simulation, we extract four spectra from different annuli with radii 30, 60, 120, and 240 arcsec (one \erosita pixel has a size of roughly 10 arcsec).
The creation of the training dataset takes several hours on 30 Intel i9 CPUs. The dataset is then randomly split into 70\% training, 15\% validation, and 15\% test data.

\paragraph{Simulation-based inference with a normalizing flow}

\begin{figure}
    \centering
    \begin{tikzpicture}
        \node (p1) at (0,0){\includegraphics[width=.4\linewidth,trim=0cm 0cm 6cm 6.5cm,clip]{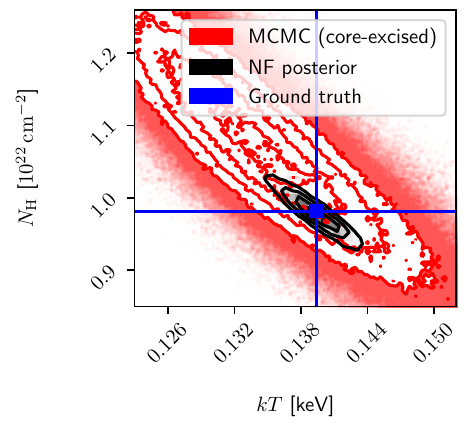}};
        \node (t1) at (p1.north){a) Pile up regime};
    \end{tikzpicture}
    \hfill
    \begin{tikzpicture}
        \node (p1) at (0,0){\includegraphics[width=.4\linewidth,trim=0cm 0cm 6cm 6.5cm,clip]{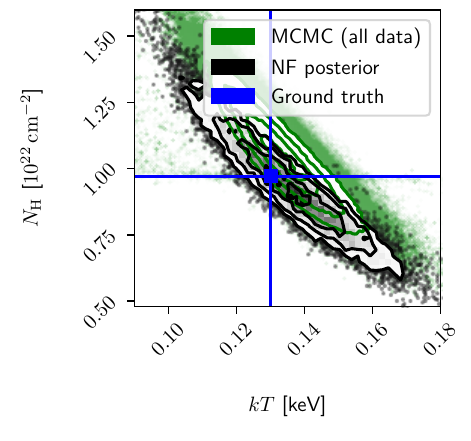}};
        \node (t1) at (p1.north){b) Non-pile up regime};
    \end{tikzpicture}
    \vspace{-.5cm} 
    \caption{Posterior distributions for two examples from the test dataset. \textit{a)} At a flux of $1.2\times 10^{-10}\,\mathrm{erg}\,\mathrm{cm}^{-2}\,\mathrm{s}^{-1}$, significant pile-up is present, and only data after excising the PSF core can be used in a traditional MCMC analysis. Red data shows the MCMC posterior from a spectrum extracted from an outer annulus with radii 120''--240'' (351 counts). Black lines depict the NF posteriors, created by sampling 10\,000 points from the distribution. Blue square denotes the ground truth. The flow posterior is more constraining than the MCMC because all data from the source region can be leveraged. \textit{b)} In the non-pile up regime, at $2\times 10^{-12}\,\mathrm{erg}\,\mathrm{cm}^{-2}\,\mathrm{s}^{-1}$, the MCMC (green) can take data from the whole source region into account (233 counts) and serve as a baseline distribution. The NF posterior is similarly sized and shaped. Contour lines denote 0.5, 1, 1.5, and $2\sigma$ confidence intervals (for a 2D Gaussian distribution).}
    \label{fig:posterior}
\end{figure}

We use a CNN with two convolutional layers, two pooling layers, and one fully-connected layer to transform the four input spectra (4$\times$1024 channels) into a lower-dimensional representation (128 nodes)\footnote{The CNN can process the broad-band behavior of the X-ray spectra well. We tested this by combining the CNN with a dense layer to directly predict point estimates for the physical parameters (Appendix Fig.~\ref{fig:testdata_parameter_estimator}).}. We use a softplus activation function which we find has better training properties than ReLU.
We then pass the lower-dimensional representation into the NF as a conditional ``context vector.'' For the NF, we use a neural spline flow \citep{durkan2019neuralsplineflows} with three transformations that morph the initial Normal distribution to the target distribution. Each transformation contains one hidden layer with 256 nodes. The output of the NF is the posterior probability distribution for the three physical parameters (flux, temperature, and absorption). We know their ``ground truth'' values from the simulation runs.
As target values for our supervised training, we standardize the ground truth values. Scaling them is particularly necessary for the flux parameter, which can vary over 4 orders of magnitude. We therefore take the logarithm of the flux before applying the standardization.
Finally, we train the CNN and the NF as one joint neural network using an Adam optimizer \citep{kingma2014adam} with a learning rate of $10^{-4}$. On 32 Intel i9 CPUs, training can be achieved in a few hours of elapsed time. No hyperparameter search is performed for this proof-of-concept demonstration.

To test our trained network, we infer parameter distributions for mock data generated using the simulator. In the pile-up regime, traditional methods usually reject data from the central source region. To compare the NF posterior for the simulated data to the PSF-core-excising method, we infer the posterior probability distribution using a Markov Chain Monte Carlo (MCMC) on the outer-most annulus spectrum (120''--240''; Fig.~\ref{fig:posterior}, \textit{right}, in red), i.e. using a likelihood without pile-up prescription. The NF reproduces the correct parameter correlation between absorption and temperature with a significantly more constrained posterior compared with the core-excising method. 
To test whether the narrow NF posterior is an artificially overconfident prediction, we also sample parameters for a data set simulated in the pile-up free (low count rate) regime with both the NF and MCMC. 
Fig.~\ref{fig:posterior} (\textit{left}) shows relatively similar distributions are obtained from both methods in the pile-up free regime. We conclude that the SBI approach has a significant advantage compared to the traditional core-excising method. As all data can be leveraged, and not only the outer region of the detector, the credible intervals are significantly smaller, leading to more precise inferences.  

\paragraph{Coverage and accuracy of the normalizing flow}

\begin{figure}
    \centering
    \includegraphics[width=1\linewidth]{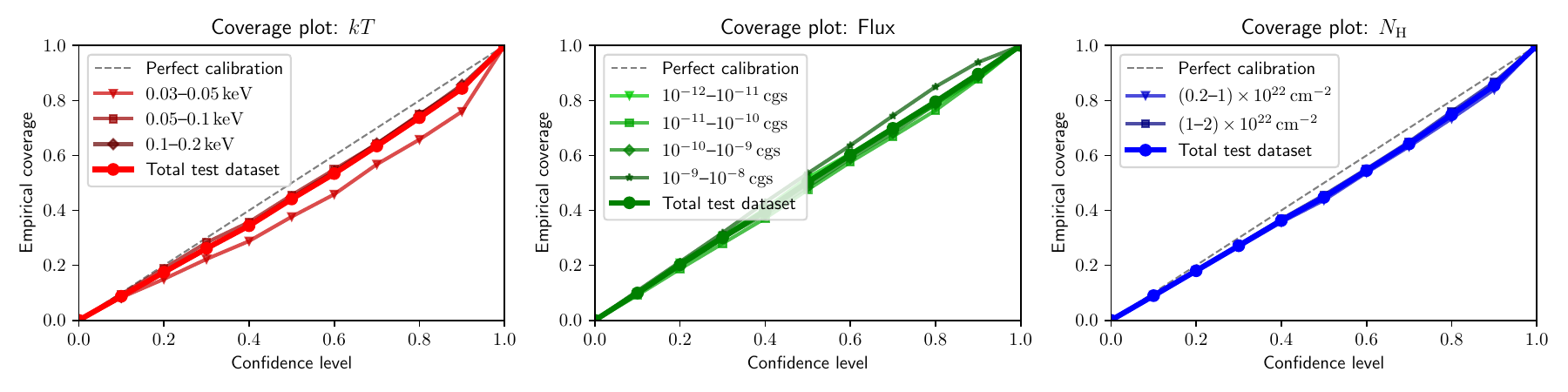}
    \vspace{-.5cm}
    \caption{Coverage plots for the three parameters of the normalizing flow. Thick lines indicates the coverage across the full test dataset (6000 sets of spectra). Thin lines denote specific ranges in the parameter space. For a perfectly-calibrated NF, the lines should converge to the bisection (gray dashed line). Regions above/below the bisection indicate under-/overconfidence, respectively.}
    \label{fig:calibration}
\end{figure}

To assess the quality of the posterior densities across the test dataset (6000 sets of 4 spatially-resolved spectra each), for each set of input spectra, we draw 10\,000 posterior samples from the NF and determine into which percentile of the distribution the ground truth falls (Eq.~7 of \citealt{Dalmasso2020a}). According to the probability integral transform (\citealt{Fisher1934}, \citealt{casella2024statistical}, p.54), if the NF is well-calibrated, the distribution of these percentiles should be uniform. 
This dependency can be visualized in so-called coverage plots \citep{deistler2022truncated}, which evaluate the measured percentiles against the expected number. As an example, if the NF is well-calibrated, any ten percentiles should contain the ground truth value 10\% of the time. 
We show the coverage of the physical parameters in Fig.~\ref{fig:calibration}. The NF is best calibrated for the flux parameter. For the absorption and temperature, which show the correlation presented in Fig.~\ref{fig:posterior}, the normalizing flow is roughly 5\% overconfident. 
We noticed a significant improvement in coverage when we doubled the training dataset from an initial 14\,000 sets of spectra to 28\,000 (the latter being the training set size for the results presented here). This can be explained by the fact that the NF needs to learn the correlation between the parameters, particularly when the posteriors are ``sharp.''
While it is reasonable to expect further improvements when increasing the training dataset and performing hyperparameter searches, the coverage indicates reasonable calibration. 

To define a realistic goal for the accuracy of the NF, we consider the systematic effects that influence our forward model. \citet{Koenig2022a} show that \texttt{SIXTE} simulations are representative of \erosita data in the pile-up regime with a systematic uncertainty of around 10\%.
We calculate the mean absolute percentage error, defined as $100\cdot 1/N \sum |y-y_\mathrm{true}|/|y_\mathrm{true}|$ where $y$ is the median of the NF posterior distribution, $y_\mathrm{true}$ is the ground truth, and the sum goes over the complete test dataset. 
We find that the mean absolute percentage error for all three parameters, particularly for the flux, is well below 10\% (see Fig.~\ref{fig:relative_bias} for details). 
This statistical accuracy is significantly smaller than the dynamic flux range that these sources can exhibit on the bright end of the source distribution (e.g., the Crab is three orders of magnitude brighter than the threshold where the \erosita detectors become piled-up; \citealt{Merloni2024a}). Constraining the parameters to within approximately 10\% is a significant improvement to rejecting the data altogether.

\paragraph{Limitations}
The current proof-of-concept neural network is trained specifically using a blackbody model, and thus expected to be biased when applied to spectra involving other physical processes.
Future work includes scaling up training to a wider range of spectral models and types of astrophysical sources.
Additional biases may arise from the pile-up implementation in the simulated training data (specifically, the charge cloud), or from the calibration of the PSF that becomes less certain at larger off-axis angles. 

\section{Conclusions}

Our SBI-based approach to inference in the presence of pile-up is motivated by the large archival datasets where pile-up has prevented a full exploration of astrophysical sources. For example, without a robust prescription for including pile-up in parameter inference, population studies are currently limited to sources below a flux threshold, above which pile-up would distort spectra too much to obtain accurate posteriors. Similarly, crucial data from bright outbursts of accreting black holes, or unexpected transients, are being discarded wholesale as unreliable.
By developing a robust inference method to assess the piled-up data, the astronomical community can leverage this underexplored discovery space.
As one salient example, the \erosita catalog contains many sources that suffer from pile-up \citep{Merloni2024a}. For example, there are approximately 36 neutron star X-ray binaries that exhibited pile-up in at least one \erosita all-sky survey (Zainab et al., in prep.). By extending the simulated training set to a wider library of spectral models, we will be able to perform efficient inference on the whole archive, flag observations in the catalog that suffer from pile-up, and provide robust posteriors for spectral parameters faster, more efficiently, more accurately, and more precisely than possible with traditional methods.

\paragraph{Acknowledgements}

We thank Rocco Di Tella for his valuable input and elaborate discussions on the calibration section of this paper. This work was conducted as part of the AstroAI collaboration.

\renewcommand{\bibsep}{0pt}
\bibliography{mnemonic,aa_abbrv,references}
\bibliographystyle{jwaabib}

\clearpage
\appendix

\renewcommand\thefigure{\thesection.\arabic{figure}}
\renewcommand\thetable{\thesection.\arabic{table}}

\section{Dataset parameter ranges}
\setcounter{table}{0}   

\begin{table}[H]
\centering
\caption{Parameter ranges of the 40\,000 simulated set of spectra. The spectral model of the astrophysical source is an absorbed black body. The training, validation, and test datasets have sizes 28\,000, 6\,000, and 6\,000, respectively. All parameters are sampled using a Latin hypercube algorithm \citep{McKayBeckmanConover1979_latinhypercube}, and the flux is sampled on a base-10 logarithmic grid.}
\label{tab:ranges}
\begin{tabular}{l|l}
Parameter   & Range \\
\hline
Flux        & $10^{-12}$--$10^{-8}\,\mathrm{erg}\,\mathrm{s}^{-1}\,\mathrm{cm}^{-2}$ \\
Temperature & $0.03$--$0.2\,\mathrm{keV}$ \\
Absorption  & $(0.2$--$2)\times 10^{22}\,\mathrm{cm}^{-2}$     
\end{tabular}
\end{table}

\section{Parameter point estimates with the CNN}
\setcounter{figure}{0}   

\begin{figure}[H] 
    \centering
    \begin{tikzpicture}
        \node (p0) at (0,0){\includegraphics[width=1\linewidth]{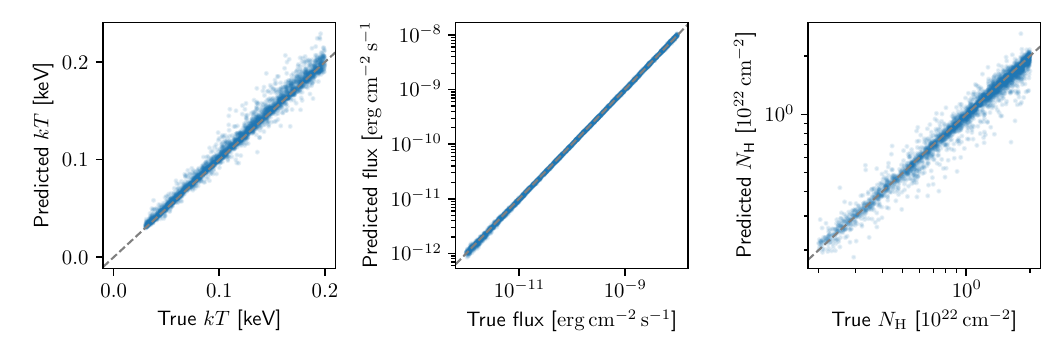}};
        \draw[red,dashed,thick] (-0.17, -1.1) -- (-0.17, 2);
        \node[red,rotate=90] (t1) at (-.4,1.05){\small low pile-up};
        \node[red,rotate=90] (t1) at (0.05,0.865){\small strong pile-up};
        \node[red] (t1) at (0.1,1.883){$\rightarrow$};
        \node[red,rotate=180] (t1) at (-.45,1.93){$\rightarrow$};
    \end{tikzpicture}
    \caption{Direct parameter reconstruction with the CNN without the normalizing flow. In this test case, we add one final fully-connected output layer of dimension three. Thus, the network directly predicts the three physical parameters. The output of all examples in the test dataset are shown. Red dashed line shows the rough threshold where \erosita data becomes piled-up \citep{Merloni2024a}, although we caution that the effect is gradual.}
    \label{fig:testdata_parameter_estimator}
\end{figure}
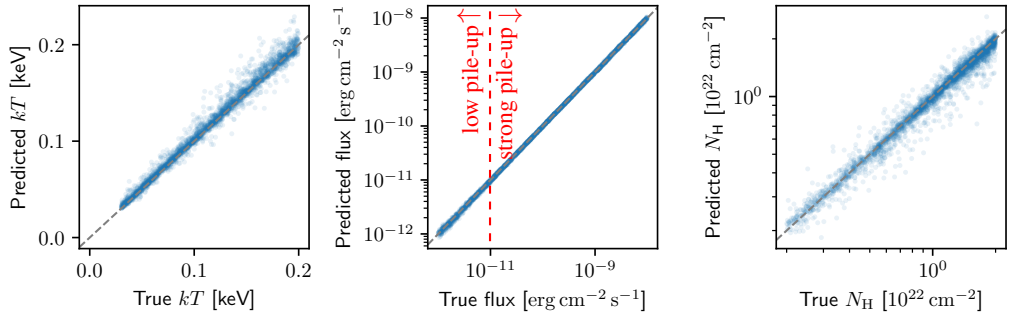

\section{Relative error}
\label{sec:app:coverage}

\begin{figure}[H]
    \centering
    \includegraphics[width=1\linewidth]{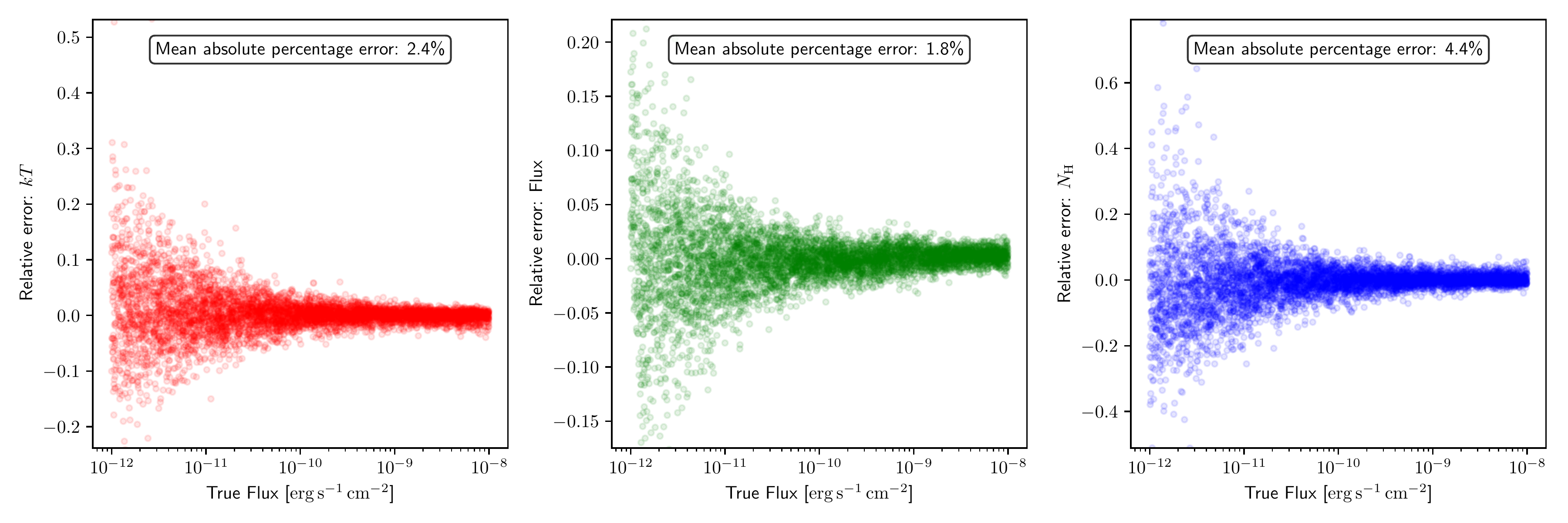}
    \caption{Relative error for the three output parameters of the normalizing flow. The relative error is defined as $(y-y_\mathrm{true})/y_\mathrm{true}$ where $y$ is the median of the NF posterior distribution and $y_\mathrm{true}$ is the ground truth. The relative error at low fluxes increases because the spectra have low number counts due to a constant simulated exposure time.}
    \label{fig:relative_bias}
\end{figure}

\end{document}